\documentclass[aps,superscriptaddress]{revtex4}
\usepackage{amsfonts}
\usepackage{amssymb}
\usepackage{amsmath}
\usepackage{youngtab}
\usepackage{theorem}

\def\DIS{\displaystyle}
\def\qed{\hfill\hbox{$\Box$}\vspace{10pt}\break}
\theoremstyle{break}
\newtheorem{Theorem}{Theorem}

\newtheorem{Proposition}{Proposition}
\newtheorem{Lemma}{Lemma}
\newtheorem{Corollary}{Corollary}

\def\ket#1{|#1\rangle}
\def\bra#1{\langle#1|}
\def\C{{\mathbb C}}
\def\R{{\mathbb R}}

\def\tr{\mbox{tr}}
\def\inf{\mbox{inf}}
\def\id{\mathbf{1}}
\def\Id{\mathbf{I}}

\begin{document}
\title{Entanglement Cost of Antisymmetric States and \\
Additivity of Capacity of Some Quantum Channels}

\author{Keiji Matsumoto}
\affiliation{
  National Institute of Informatics, \\
  2-1-2 Hitotsubashi, Chiyoda, Tokyo, 101-8430, Japan
}
\affiliation{
  Quantum computation and information project,
  ERATO, 
  Japan Science and Technology Corporation, \\
  Daini Hongo White Bldg. 201, 5-28-3 Hongo, Bunkyo, Tokyo 113-0033, Japan.
}

\author{Fumitaka Yura}
\affiliation{
  Quantum computation and information project,
  ERATO, 
  Japan Science and Technology Corporation, \\
  Daini Hongo White Bldg. 201, 5-28-3 Hongo, Bunkyo, Tokyo 113-0033, Japan.
}

\begin{abstract}
We study the entanglement cost of the states in the antisymmetric space, 
which consists of $(d-1)$ \ $d$-dimensional systems. 
The cost is always $\log_2 (d-1)$ ebits when the state is
divided into bipartite $\C^d \otimes (\C^d)^{d-2}$.
Combined with the arguments in \cite{Matsumoto02}, 
additivity of channel capacity of some quantum channels is also shown.
\end{abstract}

\maketitle

The concept of entanglement is the key for quantum communication, 
quantum computing and quantum information processing.
One candidate to quantify entanglement is entanglement of formation.
In \cite{Hayden}, it is shown that the entanglement cost $E_c$ to create
some state can be asymptotically calculated from the entanglement of
formation. In this sense, the entanglement cost has an important
 physical meaning and is significant quantity.
The known results are, nevertheless not so much
\cite{Vidal,Matsumoto02,Yura}, because it consists of minimization.
In this paper, we pay attention to antisymmetric states
that are easy to deal with.
Also, Holevo capacity of quantum channels induced by antisymmetric
spaces is discussed.

As for antisymmetric states, 
the following things are known, for example.
The entanglement of formation for two states in
  ${\cal S}\left( {\C^3}_{*} \right)$ is additive\cite{Shimono03},
  where $\C^3_* $ is defined later.
Furthermore, the lower bound to entanglement cost of density matrices
 in $d$-level antisymmetric space, obtained in \cite{Shimono02}, is
  $\log_{2} \frac{d}{d-1}$ ebit.
Recently, one of the author showed that the entanglement cost
of three-level antisymmetric states in 
${\cal S}\left( \C^3_{*} \right)$ is exactly one ebit\cite{Yura}.
In this paper, we show that the entanglement cost on
${\cal S}(\C^d_*)$ is equal to $\log_2 (d-1)$, which 
includes \cite{Yura} as a special case.

Let ${\C^d}$ be $\mbox{span}_{\C} \left\{
\ket{1}, \ket{2}, \ldots, \ket{d} \right\}$ and $d\geq 3$.
We first define the antisymmetric states which consist
of $d-1$ particles with $SU(d)$ symmetry as follows:
\begin{equation*}
  \C^d_* :=
  \mbox{span}_{\C} \left\{ \ket{1}_a, \ket{2}_a, \ldots, \ket{d}_a \right\}
  \subset {\C^d}^{\otimes \left( d-1 \right)},
\end{equation*}
where $\ket{i_1}_a := \frac{1}{\sqrt{(d-1)!}} \sum_{i_2, \cdots,i_d}\epsilon_{i_1 i_2 \ldots i_d}
  \ket{i_2} \otimes \cdots \otimes \ket{i_d} $,
  $1\leq i_1, i_2, \ldots, i_d \leq d$
  and $\epsilon$ is totally antisymmetric tensor.
When $d=4$, for example, 
$ \ket{1}_a = \left( \ket{234}-\ket{243}+\ket{342}
  -\ket{324}+\ket{423}-\ket{432} \right) / \sqrt{6}$.
Suppose $U\in SU(d)$ acts on ${\C^d}$ as $U\ket{i} = \sum_j U_{j}^{i} \ket{j}$,
then on $\C^d_*$,
\begin{align}
  U \ket{i_1}_a
&=\frac{1}{\sqrt{(d-1)!}} \sum_{i_2,\cdots,i_d}U^{\otimes (d-1)} \epsilon_{i_1 \ldots i_d} \ket{i_2 \ldots i_d} 
   =  \frac{1}{\sqrt{(d-1)!}} \sum_{j_1,\cdots,j_d}(U^{\dagger} )^{j_1}_{i_1} \epsilon_{j_1 \ldots j_d}
        \ket{j_2 \ldots j_d} \nonumber\\
  &= \sum_{j_1}(U^{\dagger} )^{j_1}_{i_1} \ket{j_1}_a,
\label{represent}
\end{align}
where we have used the fact that the totally antisymmetric tensor
  $\epsilon_{j_1 \ldots j_d}$ are invariant under $U^{\otimes d}$.
The Hermitian conjugate of $U$ in right-hand side suggests that
  $\C^d_*$ is the dual (contragredient) space of ${\C^d}$\cite{Goodman}.
The corresponding Young diagrams are
\begin{equation*}
  \mathbf{d} = \yng(1) \ , \hspace{1em}
  \left.
  \mathbf{d_{*}} = \raisebox{-5.8ex}{\yng(1,1,1,1,1)}
  \hspace*{-0.75em} \raisebox{-0.4ex}{\vdots} \hspace{0.8em}
  \right\} d-1.
\end{equation*}
Notice that the dimension of these spaces is
  $\mbox{dim} {\C^d} = \mbox{dim} \C^d_* = d$, though
  $\C^d_*$ is a multiparticle space.
Here, let us fix the space of Alice and Bob as
$\C^d_* = {\cal A} \otimes {\cal B }; \ 
  {\cal A} := {\C^d}, {\cal B}:= {\C^d}^{\otimes (d-2)}$,
and consider the entanglement between
  Alice and Bob.
The entanglement of formation $E_f$ is defined as follows:
\begin{equation}
\label{eq:ef}
E_f(\rho) = \mbox{inf} \sum_{j} p_j 
S( \mbox{tr}_{\cal B} \ket{\psi_i}\bra{\psi_i} ),
\end{equation}
where $p_j$ and $\ket{\psi_j}$ are decompositions such that 
$\rho = \sum_{j} p_j \ket{\psi_j}\bra{\psi_j}$.
Let $\Lambda_d$ be a 'partial trace channel', or
CP map from ${\cal S}(\C^d_*)$ to ${\cal S}(\C^d)$
with
$\Lambda_d(\rho)=\tr_{\cal B}\rho$.
Eq.~(\ref{represent}) implies the channel $\Lambda_d$ is contravariant,
$$\Lambda_d\left(\sum_{k,l}(U^{\dagger} )^{k}_{i} \ket{k}_a{}_a\bra{l}U^l_j\right)
=U\Lambda_d(\ket{i}_a{}_a\bra{j})U^{\dagger}$$.

Furthermore, simple calculations show,
\begin{align}
 \Lambda_d(\ket{i}_a  {}_a\bra{j})=\left\{
\begin{array}{cc}
 \frac{1}{d-1}(\id_d-\ket{i}\bra{i}) & (i=j)\\
 \frac{-1}{d-1}\ket{j}\bra{i} & (i\neq j)
\end{array}
\right. .
\label{lambda}
\end{align}
Because $\dim\C^d_*=d$, 
for any $\ket{\psi}\in\C^d_*$
there exists an element $U$ of $SU(d)$ with  
$\ket{\psi}=\sum_{k}(U^{\dagger} )^{k}_{i}\ket{k}_a$.
Hence, due to contravariancy of the channel $\Lambda_d$, 
we have
\begin{align}
 S(\Lambda_d(\ket{\psi}\bra{\psi}))=S(U\Lambda_d(\ket{i}_a{}_a\bra{i})U^{\dagger})
=S(\Lambda_d(\ket{i}_a{}_a\bra{i}))
=S\left(\frac{1}{d-1}(\id_d-\ket{i}\bra{i})\right)
=\log_2(d-1).
\end{align}
 
\begin{Proposition}
\label{prop:ef}
Let $\rho \in {\cal S}(\C^d_*)$. Then, $E_f(\rho) = \log_2 (d-1)$.
\end{Proposition}
\begin{Proof}
$
 E_f(\rho)=\inf \sum_i p_i S(\Lambda_d(\ket{\psi_i}\bra{\psi_i}))
= \inf\sum_i p_i\log_2 (d-1)=\log_2(d-1)
$
\qed
\end{Proof}

The subadditivity of $E_f$ is well known\cite{Vidal}.
\begin{align*}
E_f\left(\bigotimes_{i=1}^{n} \rho^{(i)} \right) \le
  \sum_{i=1}^{n} E_f \left( \rho^{(i)} \right), 
\end{align*}
where $\rho^{(i)}$ are density matrices on 
 ${\cal A} \otimes {\cal B}$, {\it i.e.}, bipartite states.
Using the proposition \ref{prop:ef}, we obtain the following:
\begin{Corollary}
\label{cor:sub}
For any $\rho^{(i)} \in {\cal S} (\C^{d_i}_*)$, 
$\DIS E_f \left( \otimes_{i=1}^{n} \rho^{(i)} \right) \le \sum_{i=1}^n \log_2 (d_i-1)$.
\end{Corollary}

To prove the inequality of the opposite direction, 
we use the following lemma.
\begin{Lemma}[see also \cite{Yura}]
Let $X$ be a positive semidefinite operator such that $\mbox{Tr} X=1$.
Then $\mbox{Tr} [ -X \log X] \ge - \log ( \mbox{Tr} X^2 )$.
\label{lem:yura}
\end{Lemma}
\begin{Proof}
Suppose $f(x):=-\log x$ over $\R_{+}$. It follows from the convexity of the function $f$ that
$f(\sum_i p_i x_i) \le \sum_i p_i f(x_i)$, where $\sum_i p_i = 1$, $p_i \ge 0$ and $x_i > 0$.
By setting $x_i = p_i (\forall i)$, we have
$-\sum_i x_i \log x_i \ge -\log \left( \sum_i x_i^2 \right)$.
This inequality holds even for some $x_i$ are equal to zero
  under the convention $0\log 0 = 0$.
\qed
\end{Proof}
In the followings, we denote 
the identity map from
${\cal S}({\cal K})$ to ${\cal S}({\cal K})$ by $\Id_{\cal K}$ , 
and  $\sum |X_{ij}|^2$ by $\|X\|^2$.

\begin{Lemma}
For an arbitrary state
 $\rho$ in ${\cal S}({\cal K}\otimes \C^d_*)$, we have
$
 \|\Id_{\cal K}\otimes\Lambda_d (\rho)\|^2
=\frac{1}{(d-1)^2}\left\{(d-2)\|\tr_{\C^d_*}\rho\|^2+\|\rho\|^2\right\}.
$
Here, the dimension 
 of ${\cal K}$ is arbitrary. 
\label{lem:norm1}
\end{Lemma}
\begin{Proof}
Decompose $\rho\in{\cal S}({\cal K}\otimes\C^d_*)$ into the sum
$\sum_{i,j}\ket{i}_a{}_a\bra{j}\otimes\rho_{ij}$, where
$\rho_{ij}$ are operators in ${\cal K}$.
Due to the equations $(\ref{lambda})$, we have
\begin{align*}
 \|\Id_{\cal K}\otimes\Lambda_d (\rho)\|^2
=\left\|\frac{1}{d-1}\sum_i\sum_{j\neq i}\ket{i}\bra{i} \otimes\rho_{jj}
     -\frac{1}{d-1}\sum_{i,j\neq i}\ket{i}\bra{j} \otimes\rho_{ji}\right\|^2 
=\frac{1}{(d-1)^2}\left\{\sum_k\left\|\sum_{i\neq k}\rho_{ii}\right\|^2+\sum_{i\neq j}
                                        \|\rho_{ij}\|^2\right\}
\end{align*}
The first term of the last side of the equation is rewritten as follows.
\begin{align*}
\sum_k\left\|\sum_{i\neq k}\rho_{ii}\right\|^2 
&=\sum_k\sum_{i\neq k,\, j\neq k} \tr\rho_{ii}\rho_{jj}
=(d-1)\sum_i\|\rho_{ii}\|^2+(d-2)\sum_{i\neq j}\tr\rho_{ii}\rho_{jj}\\
&=(d-2)\left\|\sum_i\rho_{ii}\right\|^2+\sum_i\|\rho_{ii}\|^2
\end{align*}
Hence, after all we have,
\begin{align*}
\|\Id_{\cal K}\otimes\Lambda_d (\rho)\|^2
= \frac{1}{(d-1)^2}\left\{(d-2)\left\|\sum_i\rho_{ii}\right\|^2
+\sum_{i,j}\|\rho_{ij}\|^2
\right\}
=\frac{1}{(d-1)^2}\left\{(d-2)\left\|\tr_{\C^d_*}\rho\right\|^2+\|\rho\|^2\right\},
\end{align*}
and the lemma is proven.
\qed
\end{Proof}

\begin{Lemma}
For any $\rho \in {\cal S} \left( {\cal K}\otimes\bigotimes_{i=1}^n{\C^{d_i}_*} \right)$, 
$\left\| \Id_{\cal K}\otimes\bigotimes_{i=1}^n\Lambda_{d_i}(\rho) \right\|^2 
\leq \prod_{i=1}^n\frac{1}{d_i-1} 
$,
where the dimension
 of ${\cal K}$ is arbitrary.
\label{lem:norm2}
\end{Lemma}
\begin{Proof}
Induction is used for the proof. First, for $n=1$, the assersion follows
 directly from lemma~\ref{lem:norm1}, because $\|\sigma\|\leq 1$ holds for any
 density matrix $\sigma$. Second, let us assume the assersion is true
 for $n-1$. Then, the lemma~\ref{lem:norm1} implies,
\begin{align*}
\left \|\Id_{\cal K}\otimes\bigotimes_{i=1}^n\Lambda_{d_i}(\rho)\right\|^2
&=\frac{1}{(d_n-1)^2}
\left\{
(d_n-2)\left\|\Id_{{\cal K}}\otimes
  \bigotimes_{i=1}^{n-1}\Lambda_{d_i}(\tr_{\C^{d_n}_*}\rho)\right\|^2
        +
\left\|\Id_{{\cal K}\otimes\C^{d_n}_*}\otimes
           \bigotimes_{i=1}^{n-1}\Lambda_{d_i}(\rho)\right\|^2
\right\} \\
&\leq
\frac{1}{(d_n-1)^2}
\left\{(d_n-2)\prod_{i=1}^{n-1}\frac{1}{d_i-1}+\prod_{i=1}^{n-1}\frac{1}{d_i-1}\right\}
=\prod_{i=1}^{n}\frac{1}{d_i-1},
\end{align*}
where the inequality in the second line comes from the assumption of
 induction.
Thus, the lemma is proven.
\qed
\end{Proof}

The following lemma is a bit weaker version of 'strong subadditivity'
\cite{Matsumoto02}.
Hereafter, 
the reduced dencity matrix 
$\tr_{\C^{d_1}_*\otimes\cdots\otimes\C^{d_{i-1}}_*\otimes\C^{d_{i+1}}_*\otimes\cdots\otimes\C^{d_n}_*}\rho$
is denoted 
by $\left.\rho\right|_{\C^{d_i}_*}$.
\begin{Proposition}
\label{prop:sup}
For any $\rho \in {\cal S} \left( \bigotimes_{i=1}^n{\C^{d_i}_*} \right)$, 
  $E_f \left( \rho \right) \ge \ \sum_{i=1}^n\log_2 (d_i-1)
=\sum_{i=1}^n E_f(\left.\rho\right|_{\C^{d_i}_*})$.
\end{Proposition}
\begin{Proof}
\begin{align*}
 E_f \left( \rho \right) 
&=\inf\sum_i p_i
 S\left(\bigotimes_{j=1}^n\Lambda_{d_j}(\ket{\psi_j}\bra{\psi_j})\right)
\geq -\inf\sum_i p_i
 \log_2\left\|\bigotimes_{j=1}^n\Lambda_{d_j}(\ket{\psi_i}\bra{\psi_i})\right\|^2\\
&\geq-\inf\sum_i p_i
 \log_2\prod_{j=1}^n\frac{1}{d_j-1}=\sum_{j=1}^n\log_2 (d_j-1).
\end{align*}
The first and the second inequality come
from lemma~\ref{lem:yura} and lemma~\ref{lem:norm2}, respectively.
 \qed
\end{Proof}
\begin{Theorem}
For any $\rho^{(i)} \in {\cal S} \left( \C^{d_i}_* \right)$, $E_f$ is
 additive, 
  $E_f \left( \otimes_{i=1}^{n} \rho^{(i)} \right) = \sum_{i=1}^n\log_2 (d_i-1)
   =\sum_{i=1}^nE_f(\rho^{(i)})$.
\label{efadditive}
\end{Theorem}
\begin{Proof}
 From the corollary \ref{cor:sub} and proposition \ref{prop:sup},
 this theorem holds.
\qed
\end{Proof}

As a corollary of this theorem, we finally obtain the first main result:
\begin{Corollary}[Main Result(1)]
$E_f \left( \rho^{\otimes n} \right) = n \log_2 (d-1)$ 
for any $\rho \in {\cal S} \left( \C^d_* \right)$.
Therefore, we obtain
\begin{equation*}
  E_c\left( \rho \right)  := 
    \lim_{n \to \infty} \frac{1}{n} E_f\left( \rho^{\otimes n} \right)
     = \log_2 (d-1).
\end{equation*}
\end{Corollary}

$E_f$ and Holovo capacity $C(\Lambda_d)$ are related with each other \cite{Matsumoto02},
\begin{align*}
 C(\Lambda_d)
:=\sup_{\{p_i,\rho_i\}}
\left\{S\left(\Lambda_d(\sum_i p_i\rho_i)\right)-\sum_i p_iS(\Lambda_d(\rho_i))\right\}
=\sup_{\rho\in S(\C^{d}_*)}\{S(\rho)-E_f(\rho)\}.
\end{align*}
Combined with 
proposition~\ref{prop:ef}, we have, 
\begin{align*}
C(\Lambda_d) =\sup_{\rho\in {\cal S}(\C^{d}_*)}S(\rho)-\log_2(d-1)
=\log_2\frac{d}{d-1}. 
\end{align*}

The following corollary, which is our second main result, is derived from
proposition~\ref{prop:sup} using almost the same argument as in the
appendix of ref.~\cite{Matsumoto02}
\begin{Corollary}
Quantum channels $\Lambda_{d_i}$ are additive,
$ C\left(\bigotimes_{i=1}^n\Lambda_{d_i}\right)
=\sum_{i=1}^n  C\left(\Lambda_{d_i}\right)=\sum_{i=1}^n\log_2 \frac{d_i}{d_i-1}
$.
\end{Corollary}
\begin{Proof}
\begin{align*}
 C\left(\bigotimes_{i=1}^n\Lambda_{d_i}\right)
&=\sup_{\rho\in S(\bigotimes_{i=1}^n\C^{d_i}_*)}\{S(\rho)-E_f(\rho)\}
\leq\sup_{\rho\in S(\bigotimes_{i=1}^n\C^{d_i}_*)}
 \left\{S(\rho)-\sum_{i=1}^n 
 E_f(\left.\rho\right|_{\C^{d_i}_*})
\right\}\\
&\leq\sup_{\rho\in S(\bigotimes_{i=1}^n\C^{d_i}_*)}
 \sum_{i=1}^n  \left\{
S(\left.\rho\right|_{\C^{d_i}_*})-
E_f(\left.\rho\right|_{\C^{d_i}_*})\right\}
\leq\sum_{i=1}^n C(\Lambda_{d_i}).
\end{align*}
Here, the first inequaltiy comes from 'strong subadditivity',
 proposition~\ref{prop:sup},
and the second inequality is due to superadditivity of joint entropy,
$ S(\rho)\leq\sum_{i=1}^nS(\left.\rho\right|_{\C^{d_i}_*})$.
Combined with well-known superadditivity of Holevo capacity 
$ C\left(\bigotimes_{i=1}^n\Lambda_{d_i}\right)
\geq\sum_{i=1}^n  C\left(\Lambda_{d_i}\right)$, the assertion is proven.
\qed
\end{Proof}

\subsubsection*{Acknowledgement}
The authors are grateful to H. Fan and T. Shimono for useful discussions
and to Prof H. Imai for support and encouragement.

\end{document}